\documentclass[preprint2,times,tighten]{aastex6}
\pdfoutput=1 
\usepackage{amsmath,amstext,natbib,float,epsfig,multirow}
\usepackage[figure,figure*]{hypcap}

\newcommand{\hMpc}{\ensuremath{h^{-1}\,\mathrm{Mpc}}}
\newcommand{\dperp}{\ensuremath{\langle d_{\perp} \rangle}}
\newcommand{\eone}{$\hat{e}_1$}
\newcommand{\etwo}{$\hat{e}_2$}
\newcommand{\ethree}{$\hat{e}_3$}
\newcommand{\mcosth}{\ensuremath{\langle \cos{\theta} \rangle}}

\shorttitle{Galaxy alignment and IGM tomography}
\shortauthors{Krolewski et al.}

\begin{document}

\title{Measuring alignments between galaxies and the cosmic web at $\lowercase{z} \sim 2-3$
using IGM tomography}
\author{Alex Krolewski\altaffilmark{1}, Khee-Gan Lee\altaffilmark{2,4}, Zarija Luki\'c\altaffilmark{2}, 
Martin White\altaffilmark{1,2,3}}
\email{krolewski@berkeley.edu}
\altaffiltext{1}{Department of Astronomy, University of California at Berkeley, New Campbell Hall, Berkeley, CA 94720, USA}
\altaffiltext{2}{Lawrence Berkeley National Lab, 1 Cyclotron Road, Berkeley, CA 94720, USA}
\altaffiltext{3}{Department of Physics, University of California
at Berkeley, Le Conte Hall, Berkeley, CA 94720, USA}
\altaffiltext{4}{Hubble Fellow}

\begin{abstract}
Many galaxy formation models predict alignments between galaxy spin and the cosmic web (i.e. the directions of filaments and sheets), 
leading to intrinsic alignment between galaxies that creates a systematic error in weak lensing measurements. These effects are often 
predicted to be stronger at high-redshifts ($z\gtrsim1$) that are inaccessible to massive galaxy surveys on foreseeable instrumentation, but 
IGM tomography of the Ly$\alpha$ forest from closely-spaced quasars and galaxies is starting to measure the $z\sim2-3$ cosmic web with the
 requisite fidelity. Using mock surveys from hydrodynamical simulations, we examine the utility of this technique, in conjunction with coeval 
 galaxy samples, to measure alignment between galaxies and the cosmic web at $z\sim2.5$.  We show that IGM tomography surveys with 
 $\lesssim5$ $h^{-1}$ Mpc sightline spacing can accurately recover the eigenvectors of the tidal tensor, which we use to define the directions 
 of the cosmic web. For galaxy spins and shapes, we use a model parametrized by the alignment strength, $\Delta\langle\cos\theta\rangle$,
 with respect to the tidal tensor eigenvectors from the underlying density field, and also consider observational effects such as errors in the
  galaxy position angle, inclination, and redshift.  Measurements using the upcoming $\sim1\,\mathrm{deg}^2$ CLAMATO tomographic survey
   and 600 coeval zCOSMOS-Deep galaxies should place $3\sigma$ limits on extreme alignment models with 
   $\Delta\langle\cos\theta\rangle\sim0.1$, but much larger surveys encompassing $>10,000$ galaxies, such as Subaru PFS, will be required
    to constrain models with $\Delta\langle\cos\theta\rangle\sim0.03$. These measurements will constrain models of galaxy-cosmic web
    alignment and test tidal torque theory at $z\sim2$, improving our understanding of the redshift dependence of galaxy-cosmic web alignment
     and the physics of intrinsic alignments.
\end{abstract}


\section{Introduction}
Gravitational collapse of the Gaussian random-phase initial conditions produced by
inflation creates a network of dense nodes connected by filaments and
sheets and separated by voids, the ``cosmic web'' \citep{zel82,kly+shand83,ein84,defw85,geller+huchra89,bond+96}.  As a result
of nonlinear structure formation,
the cosmic
web is distinctly non-Gaussian.
In the Zel'dovich approximation, collapse occurs sequentially along the
principal axes of the deformation tensor, as matter flows out of voids onto sheets,
collapses into filaments and finally streams into high-density nodes \citep{zel70}.
The accretion of matter determines the shapes and angular momenta of galaxies and their host dark matter
halos, naturally suggesting a connection between the cosmic web and galaxy
shapes and spins.



In the linear regime, the evolution of the angular momentum of a protohalo is described by tidal
torque theory \citep[TTT,][]{peeb69,doro70,whi84}.  TTT predicts that the protohalo's angular momentum will be aligned with
the intermediate eigenvector of the tidal tensor.  However, nonlinear evolution can significantly
weaken this alignment \citep{por02}, driving alignments with other preferred directions \citep[e.g. the direction of filaments, along which matter is accreting;][]{codis+12}

Alignments between the cosmic web and halo shapes and spins
have been extensively studied in N-body simulations 
\bibpunct[]{(}{)}{,}{a}{}{;}
\citep[][; see Tables 1 and 2 in
\citet{fr+14} for a recent compilation of results]{kiessling_theory_review}.
\bibpunct[; ]{(}{)}{,}{a}{}{,}
Many workers suggest that halo spins transition from parallel
to filaments at low halo mass to perpendicular to filaments at high mass
\citep{ac+07,hahn+07_evolution,codis+12,trow+13,ac14}.  
In addition,
\citet{dub+14} and \citet{codis_IA},
using the cosmological hydrodynamical simulation
HorizonAGN, argue that galaxy spin alignments exhibit a similar transition mass.
However, these results are dependent on the
measurement algorithm, simulation, and 
environmental classification \citep{kiessling_theory_review},
and it is unclear if spin-filament alignments
are the dominant spin-cosmic web alignment.
For instance,
\citet{libe+13} find a similar transition from aligned to anti-aligned
in voids and sheets as well as filaments, although with a different transition
mass in each web type, and \citet{fr+14}
find no alignment at low mass and argue that sheet alignments are as significant as filament alignments
at high mass.  In direct contradiction to the picture described above,
the cosmological zoom simulations of
\citet{hahn+10} suggest that
massive disk galaxies have spins parallel to their
host filaments while low-mass disk galaxies have spins aligned along the intermediate
axis of the tidal tensor.


In contrast, halo shape-cosmic web alignments are both stronger and more robust to measure than spin-cosmic
web alignments \citep{kiessling_theory_review},
although they have been less extensively studied.  The major axis of the halo inertia
tensor is preferentially aligned along filaments, in the plane of sheets 
\citep{alt+06,hahn+07_evolution,ac+07,zhang+09,libe+13,fr+14}
and parallel to the surface of voids
 \citep[i.e. the plane of sheets;][]{pat+06,brun+07,cuesta+08}.  Shape-cosmic
 web alignments monotonically increase
from weak to strong as a function of mass, with no transition mass 
\citep{hahn+07_evolution,ac+07,zhang+09,libe+13,fr+14}. 
Using the MassiveBlack-II cosmological hydrodynamical simulation,
\citet{chen+15} report similar results for alignments between galaxy shapes and filaments.
Galaxy shape-cosmic web alignments are closely related to ellipticity-tidal shear correlations
\citep{codis_IA}, the ``GI'' term of intrinsic alignments that is a potential major systematic 
for upcoming missions such as LSST, WFIRST
and EUCLID that aim to measure the dark energy equation of state
using weak lensing tomography \citep{hs04,bridle+07,kirk+12}.


Observational studies of galaxy-cosmic web alignment
require large numbers of galaxy redshifts to trace
the cosmic web in 3D, and are therefore primarily feasible only at low redshift. 
Observations of alignments between spiral galaxy spin and void surfaces/sheets
have produced conflicting results ranging from parallel to random to perpendicular alignment 
\citep{tru+06,sw+09,var+12}.
Locally, \citet{nav+04} find that spiral galaxy spins preferentially lie
in the supergalactic plane.
Early observations reported that spiral galaxy
spins are aligned with the intermediate axis of the tidal shear tensor
in concordance with TTT predictions \citep{lp02,le07},
and are therefore aligned perpendicular to filaments
\citep{jon+10,zhang+15}.
However, more recently \citet{temp13} and \citet{tl13}
have found that spiral galaxy spins are parallel to filaments and lenticular/elliptical galaxy
spins are perpendicular to filaments, in concordance with the transition mass
picture from simulations.  
Similarly, \citet{pah+16} find 
that elliptical galaxy spins lie perpendicular to filaments
and normal to sheets, while spiral galaxy spins exhibit much
weaker alignments along filaments.
In accordance with shape-cosmic web alignments from simulations,
\citet{zhang+13} find that galaxy major axes
preferentially align with filaments and along sheets,
a relationship that is weak for blue
galaxies and highly significant for bright red galaxies.

Similar measurements at higher redshift ($z > 0.5$) are challenging due to the difficulty of measuring
the cosmic web from the galaxy distribution, requiring large
samples of faint galaxies to achieve sufficient spatial resolution of a few Mpc,
although surveys such as VIPERS \citep{guzzo,malavasi16} are pushing this to $z\sim 0.7$.
Even with future 30m-class telescopes, it would be extremely time-consuming to obtain 
the requisite galaxy samples at $z>1$ due to the high number densities required.

At higher redshifts, Lyman-$\alpha$ forest tomography \citep{pichon2001, cau+08} 
offers an alternative method for characterizing the cosmic
web at $z \sim 2$, the epoch of peak star formation, by 
using observations of Ly$\alpha$ forest absorption in closely-spaced quasars
 and Lyman-break galaxies to reconstruct the IGM absorption field.
Using this technique, current instrumentation can probe a spatial 
resolution of a few Mpc \citep{shadow_of_colossus}, similar to the resolution of cosmic web studies
at $z<0.5$ \citep[c.f.\ the GAMA Survey;][]{eardley15}.
By simulating IGM tomographic observations with realistic signal-to-noise,
resolution, and sightline separation, we have shown that the reconstructed
flux fields visually match the underlying dark matter density \citep{lee_obs_req}
and can be used to find high-redshift protoclusters and voids \citep{stark_protoclusters,stark_voids}.
Moreover, sufficiently large surveys (with $\gtrsim 1$ deg of contiguous sky coverage)
can recover kinematically-defined cosmic web classifications with a fidelity comparable
to low-redshift surveys using the galaxy density field \citep{lee_white+16}.
These results suggest that IGM tomography could allow measurement of galaxy shape-cosmic
web
alignments at $z \sim 2.5$ in the near future.  

In this paper, we will
discuss the prospects for measuring galaxy-cosmic web
alignments using IGM tomography surveys with mean sightline separations
of $\langle d_{\perp} \rangle = [1.4,2.5,4,6.5]$ \hMpc.
These reflect both existing and possible future surveys.
Firstly, \dperp{} = 2.5 \hMpc{} reflects
the ongoing COSMOS Lyman-Alpha Mapping And Tomography Observations
(CLAMATO) survey \citep[for which the pilot phase
is being completed; see][]{lee_first_map,shadow_of_colossus}, which
aims to cover $\sim 1$ deg$^2$ in the COSMOS field using the LRIS
spectrograph on the 10.3-m Keck-I telescope.  CLAMATO will cover a redshift range
$2.2 < z < 2.5$, mapping $\sim 10^6 h^{-3}$ Mpc$^{3}$ comoving volume
with a spatial resolution of 2.5 $h^{-1}$ Mpc.
By $\sim 2020$, the Subaru Prime-Focus Spectrograph (PFS) will begin operation
\citep{takada_pfs}, and an IGM tomographic survey
building on the PFS galaxy evolution survey, but obtaining additional sightline spectra and higher
S/N,
could cover $\sim 20$ deg$^2$
with \dperp{} = 4 \hMpc{}.
On the other hand, an IGM tomography map could be constructed ``for free'' using
the $i < 24$ LBGs targeted for the PFS Galaxy Evolution Survey \citep{takada_pfs}
without additional IGM tomography targets, yielding
\dperp{} = 6.5 \hMpc{}.
Finally, the proposed FOBOS instrument on Keck will
offer much greater ($\sim 10\times$) multiplexing and field-of-view than LRIS
on the same telescope, 
allowing for deeper integrations and hence
denser sightline sampling
of $\dperp\sim 1.4$ $h^{-1}$ Mpc while surveying $\sim 1$ deg$^2$, similar to CLAMATO.

In this paper, we will estimate the quality of cosmic web direction measurements (e.g. direction of
filaments, normal vector to sheets, etc.) using mock observations based on the Nyx
hydrodynamic IGM simulations.
We will discuss the feasibility for measuring galaxy-cosmic web alignments using these
surveys in tandem with 
coeval galaxy samples at $z \sim 2.5$.

\section{Methods}
\subsection{Nyx simulations and mock observations}

We use a cosmological hydrodynamical simulation generated with the 
N-body plus Eulerian hydrodynamics $\textsc{Nyx}$
code \citep{alm_nyx}.
It has a 100 $h^{-1}$ Mpc box size with $4096^3$ cells and particles, resulting in a dark matter
particle mass of $1.02 \times 10^6 h^{-1} M_{\odot}$
and spatial resolution of 24 $h^{-1}$ kpc.  As discussed in \citet{lukic_nyx},
this resolution is sufficient to resolve the filtering scale below which
the IGM is pressure supported and to reproduce the flux statistics at percent accuracy at redshift $z=2.4$.
The box covers a similar size to the proposed CLAMATO and FOBOS survey volumes.
We use a flat $\Lambda$CDM cosmology with $\Omega_m = 0.3$, $\Omega_b = 0.047$, $h = 0.685$, $n_s = 0.965$, and $\sigma_8 = 0.8$, consistent with latest Planck measurements \citep{pl16}.

In $\textsc{Nyx}$, the baryons are modeled as an ideal gas on a uniform grid.
The baryons have a primordial composition with hydrogen and helium
mass abundances of 75\% and 25\%, respectively.  
We account for photoionization, recombination, and collisional excitation
of all neutral and ionized
species of hydrogen and helium, which evolve in ionization equilibrium with the uniform UV background given by \citet{haardt+madau12}, with the mean flux normalized to match observational values.
See \citet{lukic_nyx} for the reaction and cooling rates used in the code.
This simulation does not model star-formation and hence has no feedback from stars, galaxies, or AGNs, 
but these are expected to have a negligible effect on the Ly$\alpha$ forest statistics \citep{viel:2013}.  Future Nyx IGM simulations will include
galaxy formation physics in order to self-consistently simulate a galaxy population, 
allowing better interpretation of the relationship
between galaxies and the Ly$\alpha$ forest.

We generated $512^2$ 
absorption skewers with a spacing of 0.2 $h^{-1}$ Mpc and
sampled from these skewers to create mock data.  We computed the Ly$\alpha$
forest flux fluctuation along each skewer
\begin{equation}
\delta_F = F/\langle F \rangle - 1
\label{eqn:deltaf}
\end{equation}
where $F = \exp{(-\tau)}$ and $\tau$ is the Ly$\alpha$ optical depth,
computed in redshift space and Doppler broadened using the gas temperature.
Hereafter we refer to $\delta_F$ as the flux.

We then create mock spectra that reflect the data quality expected from 
current and upcoming surveys.  First, we randomly select absorption
skewers with the appropriate mean sightline
spacing $\langle d_{\perp} \rangle$ and rebin them along the line of sight
to a resolution of 0.78 $h^{-1}$ Mpc, similar to the line-of-sight spectral resolution from
the CLAMATO spectra.  

We simulate noise by assuming the
S/N per pixel is a unique constant for each skewer.  To determine S/N for each skewer, we draw from a power-law S/N
distribution $\textrm{d}n_{\textrm{los}}/\textrm{dS/N}
\propto \textrm{S/N}^{-\alpha}$ \citep[][hereafter S15b]{stark_protoclusters},
where S/N ranges between 1.5 \citep[the minimum S/N in CLAMATO;][]{lee_first_map,shadow_of_colossus} and infinity.
\citet{lee_obs_req} found that $\alpha \sim 2.5$ for the LBGs and QSOs that we 
target; however, as the sightline separation increases, the sources targeted become brighter
and we move further along the exponential tail of the luminosity function, so
$\alpha$ becomes larger.  S15b find $\alpha = 2.9$ (3.6) for
$\langle d_{\perp} \rangle = 2.5$ (4) $h^{-1}$ Mpc, respectively.  
They did not determine $\alpha$
for sightline spacings $< 2$ \hMpc{} or $> 4$ \hMpc; therefore
we use $\alpha = 2.7$ (3.6) for our \dperp{} = 1.4 (6.5) \hMpc{} map
by extrapolating the S15b values for \dperp{} = 2 (4) \hMpc{}.
We confirm the power-law distribution is appropriate
by comparing it to the S/N distribution of observed pixels from the CLAMATO pilot observations.
Using the simulated S/N
distribution, we add noise to each pixel assuming a Gaussian distribution.  We also
model the effect of continuum-fitting error with an RMS of 10\%:
$F_{\textrm{obs}} = F_{\textrm{sim}}/(1+\delta_{\textrm{cont}})$ where 
$\delta_{\textrm{cont}}$ is a random Gaussian
deviate, identical for all pixels within a skewer, with mean 0 and standard deviation 0.1.
This reflects the continuum-fitting uncertainties expected from data with comparable
S/N  \citep{lee+12}.

For the tomographic reconstruction, 
we use the publicly available Wiener filter reconstruction code of \citet{stark_protoclusters}
\footnote{\url{https://github.com/caseywstark/dachshund}}
to create a 3D map of the flux field.  The Wiener filter is ideal for reconstruction
as it provides a minimum variance estimate of the 3D field, assuming the field is normally
distributed \citep{cau+08,lee_obs_req,stark_protoclusters}.  The reconstructed signal
is
\begin{equation}
\hat{s} = \textbf{S}_{\textrm{md}}(\textbf{S}_{\textrm{dd}} + \textbf{N})^{-1} \textbf{\textit{d}}
\label{eqn:wiener}
\end{equation}
where $\textbf{\textit{d}}$ is the data, $\textbf{N}$ is the noise covariance,
$\textbf{S}_{\textrm{md}}$ is the map-data covariance, and $\textbf{S}_{\textrm{dd}}$ is
the data-data covariance.  We assume that the noise covariance is diagonal, so that $N_{ij} =
n_i^2 \delta_{ij}$ where $n_i$ is the simulated noise for each pixel.  We further assume
that $\textbf{S}_{\textrm{md}} = \textbf{S}_{\textrm{dd}} = S$:
\begin{equation}
S = \sigma_F^2 \exp{\left[-\frac{\Delta x_{\perp}^2}{2 l_{\perp}^2}
- \frac{\Delta x_{\parallel}^2}{2 l_{\parallel}^2}
\right]}
\label{eqn:covariance}
\end{equation}
We use $\sigma_F^2 = 0.05$
and isotropic smoothing with $l_{\parallel} = l_{\perp} = 
\langle d_{\perp} \rangle$.
Hereafter we refer to the reconstructed flux as $\delta_F^{\textrm{rec}}$ and
the simulated flux as $\delta_F$.


\subsection{Defining the Cosmic Web}

We measure the cosmic web directions using
the eigenvectors of the deformation tensor,
an approach inspired by the cosmic web classifications
of \citet{hahn+07_prop}.
While there are many alternative cosmic web classifiers
\citep[see enumeration in][]{caut14}, we prefer the deformation
tensor approach for a variety of reasons: it allows direct comparison with \citet{lee_white+16}; it is physically motivated by the Zel'dovich approximation;
and it is directly related to the gravitational shear field relevant for weak-lensing
intrinsic alignments \citep{codis_IA}.

The deformation tensor is defined as the Hessian of the gravitational
potential $\Phi$:
\begin{equation}
D_{ij} = \frac{\partial^2 \Phi}{\partial x_i \partial x_j}
\label{eqn:hessian}
\end{equation}
The Hessian is most efficiently calculated in Fourier space,
using the Poisson equation in suitable units where $4 \pi G = 1$,
$\nabla^2 \Phi = k^2 \Phi = \delta_k$.  Therefore
we can directly compute the Fourier transform of $\widetilde{D}_{ij}$
from the density:
\begin{equation}
\widetilde{D}_{ij}(k) = \frac{k_i k_j}{k^2} \delta_k
\label{eqn:def_tensor_kspace}
\end{equation}
and inverse-Fourier transform this quantity to obtain $D_{ij}$.
To compute $D_{ij}$, we define $\delta$ as the sum
of the matter and baryonic overdensity measured in redshift space,
binned on a $128^3$ grid and smoothed with a Gaussian kernel with
standard deviation $R_G = 2$ \hMpc{} (see below for further description of smoothing).
We then diagonalize the deformation tensor at every point in space
to obtain its eigenvectors, $\hat{e}_1$, $\hat{e}_2$, and $\hat{e}_3$
(where the eigenvectors correspond to eigenvalues
$\lambda_1 < \lambda_2 < \lambda_3$; in the Zel'dovich approximation,
collapse proceeds first along \ethree{} and last along
\eone{}). Note that the traceless tidal shear tensor
\begin{equation}
T_{ij} = \frac{\partial^2 \Phi}{\partial x_i \partial x_j} - \frac{1}{3} \nabla^2 \Phi \delta_{ij}
\end{equation}
which is more relevant than $D_{ij}$ for intrinsic alignment,
shares its eigenvectors with the deformation tensor.  As a result, we will use the phrases
``eigenvectors of the tidal tensor'' and ``eigenvectors of the deformation tensor''
interchangeably.

We define the cosmic web directions of the IGM tomography map,
which reconstructs the Ly$\alpha$ forest flux,
as the eigenvectors of the pseudo-deformation tensor \citep{lee_white+16},
where we simply substitute the Fourier transform of the flux field, $\delta_F$,
for $\delta_k$ in Equation~\ref{eqn:def_tensor_kspace}.
Since $\delta_F$ has the opposite sign as
$\delta_k$, we order the eigenvalues of the pseudo-deformation tensor
from largest to smallest.

We classify each point as a node, filament, sheet or void
using the number of eigenvalues greater than a 
nonzero threshold value $\lambda_{\textrm{th}}$, similar
to \citet{lee_white+16} and \citet{fr+09}.
A nonzero threshold leads to a better agreement
with visually prominent sheets and filaments \citep{fr+09}
and is justified because
directions with a small positive $\lambda$ are contracting so slowly
they may not collapse in a Hubble time.
Similar to \citet{lee_white+16}, we choose $\lambda_{\textrm{th,m}}$ by matching the volumetric void fraction in the matter density to the
$\sim 19\%$ void fraction of \citet{stark_voids}.
We choose the threshold for the flux,
$\lambda_{\textrm{th,F}}$, using the same condition on the
void fraction for the \dperp{} = 1.4, 2.5, 4, and 6.5 \hMpc{} reconstructions.

\begin{figure}[h]
\psfig{file=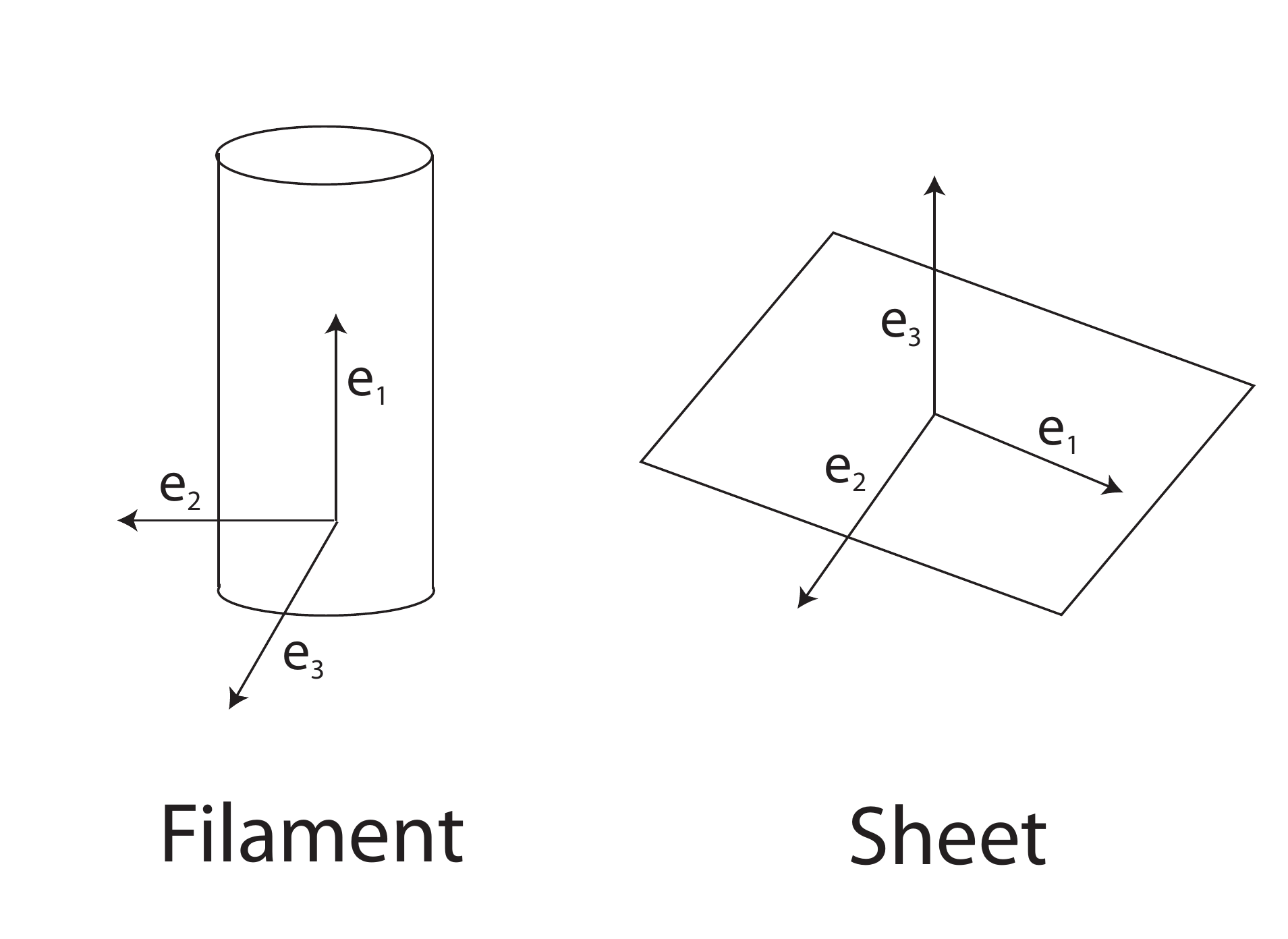,width=8 cm,clip=}
\caption[]{\small
Relationship between eigenvectors of the tidal tensor (or equivalently, deformation tensor) and cosmic web directions.}
\label{fig:eigenvectors_cosmic_web}
\end{figure}

\begin{figure*}
\hspace*{-1cm}
\includegraphics[width=20cm]{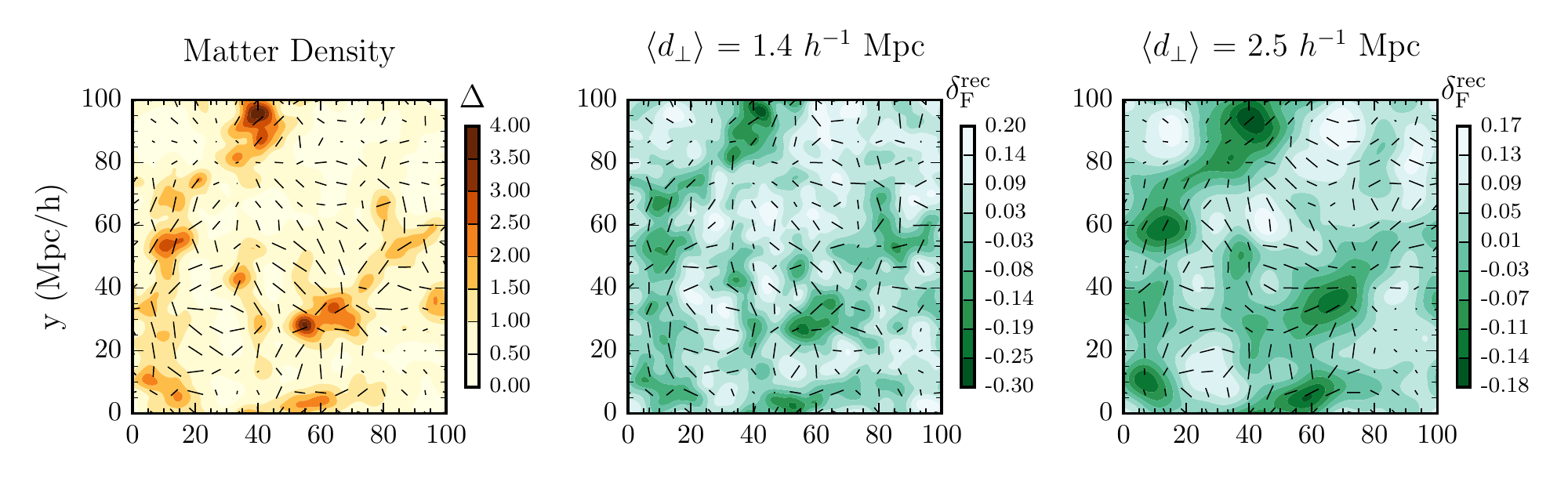}
\hspace*{-1cm}
\vspace*{-0.3cm}
\includegraphics[width=20cm]{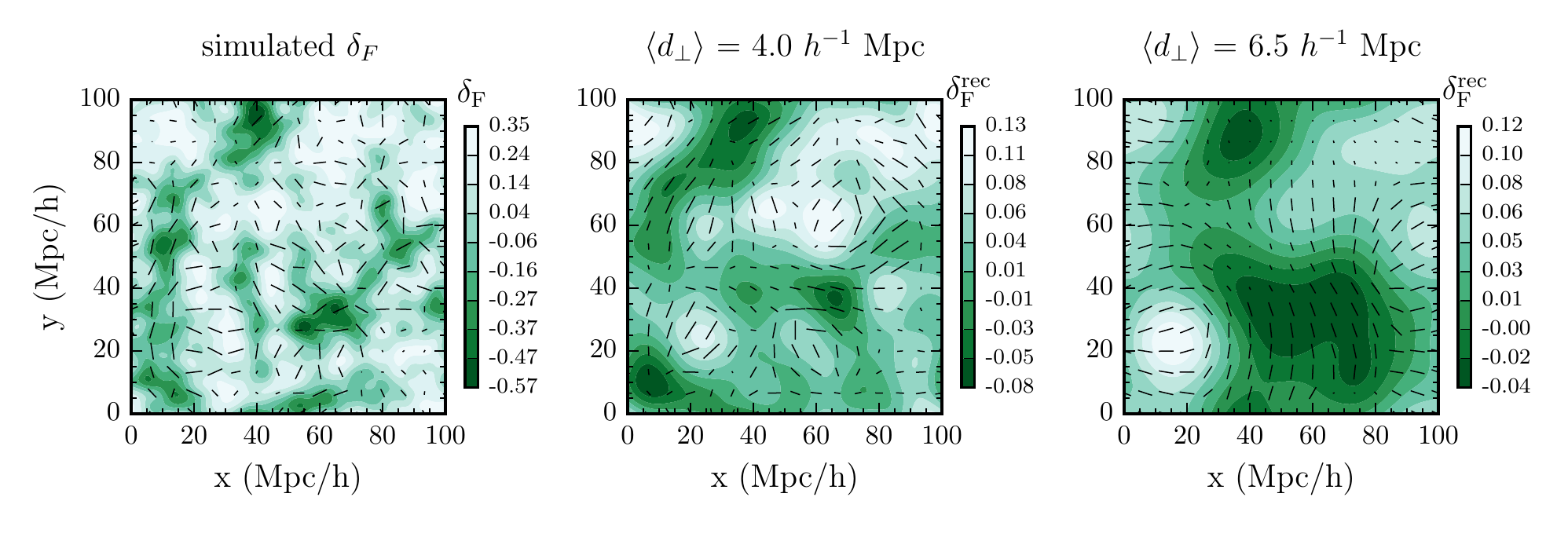}
\caption[]{\small
Matter density (dark matter plus baryons) and $\delta_F$ from the simulations (left) and 
$\delta_F^{\textrm{rec}}$ from simulated IGM tomography observations with varying
\dperp{} (right).  All quantities are evaluated in redshift space.
The figures show a slice through the full
simulation box with width 0.8 \hMpc{} along the line of sight. The overplotted vectors are
 \eone determined from the corresponding field.}
\label{fig:density_fields_e1}
\end{figure*}

The eigenvectors of the deformation tensor
are related to the underlying geometry and kinematics
of the cosmic web.  In the Zel'dovich approximation, matter is collapsing along the eigenvector
when the eigenvalue is positive, and expanding
along the eigenvector when the eigenvalue is negative \citep{hahn+07_prop}.
In a filament, there is only one negative
eigenvalue, thus $\hat{e}_1$ is the only direction
of expansion, making it the direction along which the filament
is oriented.  Similarly, in a sheet, there is only one positive
eigenvalue, making $\hat{e}_3$ the normal vector to the sheet.
Therefore, we define the directions
of the cosmic web using the eigenvectors of the tidal tensor.

Following \citet{lee_white+16},
we smooth the deformation tensor to minimize the effects of reconstruction
noise and remove small-scale fluctuations.  They found that a
Gaussian kernel with $R_G \sim 1.5 \langle d_{\perp} \rangle$
was appropriate for this purpose (see also \citealt{cau+08}).  Therefore, we use
smoothing kernels with $R_G = [2, 4, 6, 10]$ for \dperp{} = [1.4, 2, 4, 6.5]
reconstructions.
A larger smoothing scale leads to a more homogeneous map with less
variation in $\delta_F^{\textrm{rec}}$, so maps with
a larger \dperp{} have a smaller spread in $\delta_F^{\textrm{rec}}$
(Figure~\ref{fig:density_fields_e1}).

We also smooth the underlying matter density which we use for comparison.  
Since our cosmic web classification scheme is based on the Zel'dovich
approximation, it is only valid up to the mildly nonlinear
scales where the Zel'dovich approximation fails \citep{eardley15}.  Therefore,
smoothing on scales of a few \hMpc{} is appropriate to eliminate
highly non-linear scales.  We choose a matter smoothing
scale of 2 \hMpc, comparable to the smoothing
scales used in shape-cosmic web and spin-cosmic
web alignment studies from simulations \citep[see compilation in][]{fr+14}.

The choice of smoothing scale is ultimately arbitrary; for instance, we could
follow \citet{lee_white+16} and choose a different matter density smoothing
scale for each reconstruction, matching the smoothing scales of the flux
map and the matter.  
However, we prefer to use a single ``true'' matter map to generate galaxy
spins (see Section~\ref{sec:align_model}).  
The results are qualitatively similar if we instead match the matter smoothing
scale to the reconstruction smoothing scale, in that the recovery of the eigenvectors
declines from \dperp{} = 1.4 \hMpc{} to 6.5 \hMpc{}.  However, the decline is much
less steep if we match smoothing scales; thus, much of the misalignment between
the matter field and IGM tomography maps with \dperp{} $> 2$ \hMpc{} is due to the
mismatch in smoothing scales.

Figure~\ref{fig:density_fields_e1} compares the matter density field
to the reconstructed flux field for \dperp{} = 1.4, 2.5, 4, and 6.5 \hMpc{}
for an 0.8 \hMpc{} slice through the simulation box.  
We also show the simulated
redshift-space $\delta_F$ field, equivalent to a ``perfect resolution'' IGM tomography
survey. 
We publicly release the matter field, simulated $\delta_F$,
IGM tomography maps, and halo catalogs\footnote{\url{http://tinyurl.com/hg7u4dg}}.
 We overplot headless vectors corresponding to the projection of \eone{}
onto the $xy$ plane; arrowheads are not displayed because the sign of \eone{}
is arbitrary.
The IGM tomography surveys are smoothed using the smoothing scales
defined above, while the matter density and simulated flux field, $\delta_F$, are smoothed
with $R_G = 2$ \hMpc{}.  The simulated flux and \dperp{} = 1.4 \hMpc{} fields
reproduce the matter density well, although with a very different
dynamical range owing to the nonlinear transformation from $\delta$ to $\delta_F$.
The simulated flux and \dperp{} = 1.4 \hMpc{} maps reproduce the small-scale
structure in the matter-density \eone{} quite well, while larger smoothing scales
yield a smoother distribution of $\delta_F^{\textrm{rec}}$ and \eone{} that captures the large-scale
features but misses much of the smaller-scale structure.

\vspace{20pt}

\subsection{Galaxy spin observations}

Measuring galaxy-cosmic web alignments at $z \sim 2.5$
requires a large galaxy sample with accurate redshifts and
structural parameters.  The typical half-light radius of
a $z \sim 2.4$ galaxy is $\sim 0.3$'' \citep{giavalisco1996,vdw+14_3dhst}.
Space-based or adaptive-optics observations are therefore
preferred for measuring structural parameters at $z \sim 2.4$,
although the shapes of faint galaxies can be measured well from
the ground with deep exposures and $\sim 0.5$'' seeing
\citep{chang+13}.  Additionally, LSST plans to measure cosmic
shear out to $z \sim 3$ using ground-based observations with
0.7'' mean seeing \citep{lsst_book}.

Galaxy spins can be determined from galaxy images using
the galaxy's position angle and axis ratio assuming
the galaxy is an oblate spheroid \citep{hg84}. The position angle defines the
direction of the projected major axis, while the inclination is
calculated from the axis ratio, assuming the galaxy is circular if viewed
face-on and the intrinsic thickness is known.
As a result, this method is only applicable
to spiral or spheroidal galaxies, which can be approximated as
oblate spheroids \citep[e.g][]{le07,temp13}.
For elliptical galaxies, alignments between the galaxy's major axis and the projected
eigenvectors of the tidal tensor are most relevant.

The spin axis is then assumed to be the minor axis of the galaxy ellipsoid.
In principle the spin and minor axis may be misaligned,
though both observations and simulations find small misaligments \citep{fran91,codis_IA},
$\sim 15^{\circ}$
at $z \sim 2$ \citep{wis+15}.
High resolution or adaptive-optics IFU observations
can measure the kinematic major axes of galaxies at $z \sim 2$
\citep[e.g.][]{wis+15}, but even large surveys
with KMOS or NIRSpec \citep{giar+16}
may only accumulate 1000 galaxies over the next several
years, far smaller than the sample sizes expected from wide-field imaging
surveys.

Galaxies at $z \sim 2$ have a clumpier and more irregular
morphology
than galaxies at low redshift \citep{lotz06}.
This may present an additional challenge for measuring the major axis
and ellipticity of the light profile.  In addition, the intrinsic
thickness is a major source of systematic error in determining
the inclination: 
it varies by morphology \citep{hg84}
and potentially also with redshift, 
as galaxies at $z \sim 2$ are thicker than galaxies
at low redshift \citep{law12}. Furthermore, only the absolute
value of the inclination is measurable, so 
there will always be a degeneracy between spins pointing towards the observer and spins pointing away
from the observer, except for face-on or edge-on galaxies.
Alignment studies have attempted to mitigate this
degeneracy in several ways \citep{le07,var+12,sw+09,pah+16,tru+06}
but ultimately the degeneracy will degrade the measured
alignment signal.

To understand the impact of these systematics on alignment
measurements at $z \sim 2$, we include realistic errors
in the galaxies' position angles and inclinations.
Since the inclination measurements may be particularly
impacted by systematic errors from the unknown
intrinsic thickness and anisotropies
in the PSF, we consider both 3D and 2D alignment measurements,
where the 2D alignment measurements ignore the $z$-direction
of the eigenvectors altogether.  To model the spin degeneracy,
we randomly select each spin to face either towards or away
from the observer.

\subsection{Galaxy alignment model}
\label{sec:align_model}

We wish to remain agnostic about the mechanisms
and strength of the galaxy-cosmic web alignment signal.
Hence, we assign galaxy spins to the simulation halos
using a stochastic relationship between the eigenvectors
of the tidal tensor of the matter density field
and galaxy spin. This prescription allows us greater
flexibility to adjust the strengths of galaxy alignments
compared to adhering to the 
results of a single simulation. The galaxy formation
physics governing the shape-cosmic web 
relationship at $z \sim 2$ are not well understood,
so flexibility in modeling this relationship is valuable.
Moreover, it is not clear which eigenvector the spin
is most strongly aligned with, so our model allows us to adjust
the alignment strength with any of the three eigenvectors.

Our model takes as input $\langle \cos{\theta} \rangle$,
the mean of the cosine of the angle between
the galaxy spins and the local eigenvector
of the matter field tidal tensor.
We parametrize the PDF of $\mu \equiv \cos{\theta}$ as
\begin{equation}
P(\mu) = a\mu^2 + c
\label{eqn:align_pdf}
\end{equation}
where $c = 1-a/3$ such that $P(\mu)$ is normalized.
For simplicity, we consider only the alignment with a single
eigenvector at a time.

To compare to various observational and simulation
results,
we parameterize $P(\mu)$ using $\langle \cos{\theta} \rangle$
rather than $a$:
\begin{equation}
a = 12 \langle \cos{\theta} \rangle - 6
\label{eqn:a_to_mean}
\end{equation}
Equation~\ref{eqn:align_pdf} roughly reproduces
$P(\mu)$ as measured from simulations.
A representative value of \mcosth{} is given by 
the simulations of \citet{codis_IA},
\mcosth{} = 0.509.  This value is similar to \mcosth{} for \eone{}-spin alignments
measured
at low redshift \citep{tl13,zhang+13,pah+16} and spin-filament alignments
in simulations \citep{dub+14} and observations \citep{temp13}.
Moreover, for small
deviations from random alignments, Equation~\ref{eqn:align_pdf}
agrees well with Equation 5 in \citet{le07},
who derive an analytic expression for the misalignment
angle between galaxy spin and \etwo{} in tidal torque theory.



To assign each galaxy a spin axis, we start by 
assigning a redshift-space error drawn
from a normal distribution with standard deviation
$\sigma_v$ (see Section~\ref{sec:survey_params}).  Next, we
find
the eigenvectors of the tidal tensor of the matter density field
at the nearest grid point to the galaxy's redshift-space position,
including redshift error.
The misalignment angle $\theta$ is randomly drawn from
Equation~\ref{eqn:align_pdf} and the azimuthal
angle $\phi$ from the uniform distribution between 0 and $2\pi$.
Since both the eigenvectors and the galaxy spin axis are headless vectors,
we randomly generate a direction for the galaxy spin as well.
Last, we add a random Gaussian deviate with standard deviation
$\sigma_{\textrm{PA}}$ ($\sigma_i$) to the position angle (inclination),
and randomly choose whether the galaxy spin will be oriented
towards or away from the observer.

\subsection{IGM and galaxy survey parameters}
\label{sec:survey_params}

Our fiducial parameters are sightline spacing of
2.5 \hMpc{}, coeval sample of 10000 galaxies, redshift
errors of 100 km s$^{-1}$, $\sigma_{\textrm{PA}} = 10^{\circ}$
and $\sigma_i = 10^{\circ}$.

We consider how the measurable significance of the alignment
signal varies with different sightline spacings, galaxy sample
sizes and assumed errors.  We use sightline spacings of
 6.5, 4.0, 2.5, and 1.4 \hMpc{}, as well
 reconstructions of the full noiseless $\delta_F$ 
 simulation grid (with 0.8 \hMpc{} voxels)
as the limiting case of ``perfect'' IGM tomography.
The 4.0, 2.5, and 1.4 \hMpc{} spacings correspond to the
sightline spacings expected for the PFS, CLAMATO, and FOBOS
IGM tomography surveys, respectively, while 6.5 \hMpc{}
is the sightline spacing of an IGM reconstructions using only
the baseline PFS galaxy evolution survey \citep{takada_pfs} 
without incorporating additional targets for tomography.

We use coeval galaxy samples of 600, 3000, 10000 and 30000 galaxies.
The density of target galaxies for tomography differs by an
order of magnitude between the \dperp{} = 1.4 and 
 4.0 \hMpc{} cases, so the galaxy samples were chosen
 to span an order of magnitude as well, allowing
 us to directly compare the importance of coeval galaxy sample size
versus sightline density.
The fiducial 10000 galaxy sample does not require 10000 galaxies in 1 deg$^2$ (the angular
size of our simulation box at $z \sim 2.4$); rather, the galaxies may be spread over a wider area
if the tomographic map also has the same coverage.  The cosmic
web recovery does not depend on halo mass of the galaxies (Figure~\ref{fig:e1_by_mass}),
so we emulate a survey with larger area by simply
including lower-mass galaxies in our sample.

The 10000 galaxy sample is similar
to the number of $2.15 < z < 2.55$ redshifts that
the PFS galaxy evolution survey could obtain. Structural
parameters for such a sample could be measured either
from the deep HyperSuprimeCam imaging used for PFS
targeting or from wide-field space-based imaging
from Euclid or WFIRST.  More ambitious upcoming
surveys could obtain even larger samples of coeval
galaxies, such as the proposed ``Billion-Object Apparatus'',
which could measure redshifts for $10^5$ coeval galaxies
per square degree in the 2030s \citep{dod+16} --- 
as a conservative choice, we therefore include a 30000 galaxy
sample to represent these futuristic surveys.
At the other extreme, we also consider a 600 galaxy sample, roughly
matching the number of coeval galaxies in the CLAMATO
volume, primarily from the zCOSMOS-deep survey
 \citep{scoville:2007,liil+07}.



The fiducial redshift
errors are 100 km s$^{-1}$, appropriate for redshifts
from nebular emission lines in restframe optical spectra \citep{steid+10}.  We consider redshift errors of 300 km s$^{-1}$,
appropriate for redshifts from UV absorption lines or Ly$\alpha$
emission lines \citep{steid+10,kriek_mosdef}, and
500 km s$^{-1}$, appropriate for emission-line redshifts
from grism spectra \citep{kriek_mosdef,momcheva2016}.
We also consider
the maximal redshift errors allowed by our box size,
in which the $z$ position of each galaxy in the box
is drawn from a uniform distribution.  This produces
a distribution with $\sigma_v \sim 2000$ km s$^{-1}$,
somewhat better than typical photometric redshifts 
($\sigma_v \gtrsim 9000$ km s$^{-1}$)
or Ly$\alpha$ tomographic redshifts
\citep[$\sigma_v \gtrsim 3000$ km s$^{-1}$;][]{sch+16}.

The fiducial error on the galaxy position angle is 10$^{\circ}$,
consistent
with position angle errors as estimated from both
cosmic shear measurements from HST imaging \citep{leau+07,joa+13_1}
and from structural parameter measurements using CANDELS
imaging \citep{vdw+12}\footnote{
These methods differ most importantly in that the weak-lensing
analyses are somewhat more careful about accounting
for systematic errors from PSF variation than the galaxy shape
analyses.  Also, the weak-lensing analyses present
their results in terms of error on the galaxy polarization
or ellipticity, while the galaxy shape analyses directly
report the errors on the position angle.  Polarization/ellipticity
errors can be translated to PA errors using Taylor series error propagation.} for galaxies with magnitudes, sizes, and Sersic
indices typical of $z \sim 2$ galaxies.
We also consider position angle errors of 5$^{\circ}$
and 20$^{\circ}$ in order to determine the importance
 of imaging quality.
 These position angle errors may not be appropriate for ground-based
 imaging, which generally suffers from increased uncertainty in shape modeling
 \citep[e.g.][]{chang+13}.  We therefore also consider position angle
 errors of 40$^{\circ}$, which may be more realistic for structural
 parameters derived from ground-based imaging.
 
We use a fiducial inclination error of 10$^{\circ}$.
The error on the inclination can be related to the error
on the ellipticity using
Taylor series error propagation.  Using the ellipticity
errors from the CANDELS catalog for $z \sim 2$ galaxies
assuming intrinsic thickness 0.25 \citep{vdw+12}, we find a median $\sigma_i = 6^{\circ}$.
To conservatively account for systematic errors from the intrinsic
thickness, the fiducial value of $\sigma_i$ is $10^{\circ}$.
We also consider $\sigma_i = 5^{\circ}$
and $\sigma_i = 20^{\circ}$ to determine the impact
of inclination error on our measurement.

For each measurement, we simulate galaxy selection
by randomly selecting $N_{\textrm{gal}}$ halos with
$M_h > 10^{10.5} M_{\odot}$.  While this is does not reflect
a realistic selection function, the
cosmic web recovery does not depend on mass
(Figure~\ref{fig:e1_by_mass})
so we expect similar results for realistic selection
functions.  This also allows us to mock up larger-area
surveys without using a larger simulation box, as we can
simply select more galaxies within the same 100 \hMpc{}
volume.

To simulate the cosmic web-galaxy spin alignment measurement,
we measure $\langle \cos{\theta_{rg}} \rangle$, the dot product
of 
the reconstructed eigenvector and the galaxy spin,
as a function of $\langle \cos{\theta_{eg}} \rangle$, where $\theta_{eg}$
is the angle between the matter field eigenvector and the galaxy spin.
We turn this into a significance above random by subtracting 0.5,
the mean of $\cos{\theta}$ for a random distribution, and dividing
by the standard deviation of 1000 simulations of the measurement.
For 2D measurements, the significance is defined as $\langle {\theta_{rg}} \rangle - \pi/4$
divided by the standard deviation, since a random vector in 2D follows
a uniform distribution in $\theta$.

\section{Results}

\subsection{Recovery of cosmic web directions}

We first compare our cosmic web classification to
\citet{lee_white+16}, who use an N-body simulation rather than
a hydrodynamic IGM simulation and match the DM and IGM tomography smoothing scales.  We confirm their finding
that IGM tomography can recover cosmic web classifications
with similar fidelity to low-redshift surveys, suggesting
this finding is insensitive to the details of the 
simulation and the choice of smoothing scale.


The fraction of the volume with $\Delta N ^+$ = 0
is somewhat lower in our $\langle d_{\perp} \rangle$ = 2.5 \hMpc{}
reconstruction
than that of 
\citet{lee_white+16}, as they find [15,69,15]\% of the volume was classified
within [-1,0,+1] eigenvalues.  However, we do not match
the matter smoothing scale to the IGM tomography smoothing scale
and we include continuum errors in our mock absorption skewers, both of which
degrade the reconstructions.
Compared to \citet{lee_white+16},
we recover sheets, voids, and filaments with slightly lower fidelity,
and nodes with significantly lower fidelity.

\begin{deluxetable*}{c c c  c c c c c c c}
\tablecolumns{10}
\tablecaption{\label{tab:cosmic_web_classification} Fidelity of Cosmic Web Classification}
\tablehead{
\multirow{2}{*}{$\langle d_{\perp} \rangle$} & 
\multirow{2}{*}{Smoothing} & 
\multirow{2}{*}{Flux Eigenvalue} &
\multicolumn{3}{c}{Fraction by} &
\multicolumn{4}{c}{\multirow{2}{*}{Volume overlap (\%)}}
\\
& & & &  \multicolumn{3}{c}{$\Delta N^{+}$ (\%)} 
\\
(\hMpc) & (\hMpc) & Threshold & \colhead{-1} & \colhead{0} & \colhead{1}
& \colhead{Node} & \colhead{Filament} & \colhead{Sheet} & \colhead{Void}
}
\startdata
$\delta_F$ & 2 & $\lambda_{\textrm{th,F}} < -0.0175$ & 8.8 & 83.9 & 7.2 & 67.0 & 82.9 & 85.2 & 84.7 \\
1.4 & 2  & $\lambda_{\textrm{th,F}} < -0.0101$ & 15.2 & 69.2 & 14.7 & 45.7 & 66.1 & 72.8 & 67.5 \\
2.5 & 4  & $\lambda_{\textrm{th,F}} < -0.0095$ & 17.9 & 61.1 & 18.9 & 24.0 & 54.2 & 67.3 & 60.2 \\
4.0 & 6 & $\lambda_{\textrm{th,F}} < -0.0090$ & 22.4 & 52.5 & 20.7 & 16.7 & 46.8 & 59.3 & 48.3 \\
\enddata
\tablecomments{$\Delta N^+ = N^+_{\textrm{matter}} - N^+_{\textrm{F}}$
where $N^+$ is the number of eigenvalues with $\lambda > \lambda_{\textrm{th}}$ in a given map.
Fraction by $\Delta N^+$ refers to the volume fraction of the map where $\Delta N^+$
has that value.
Volume overlap is the fraction of all points classified as a particular web
element in the matter field that are also classified as that web element in the flux
map.  We use $\lambda_{\textrm{th,m}}$ = 0.043 and a smoothing scale of 2 \hMpc{} for the matter field.}
\end{deluxetable*}



\begin{figure*}
\hspace*{-1cm}
\centering{\psfig{file=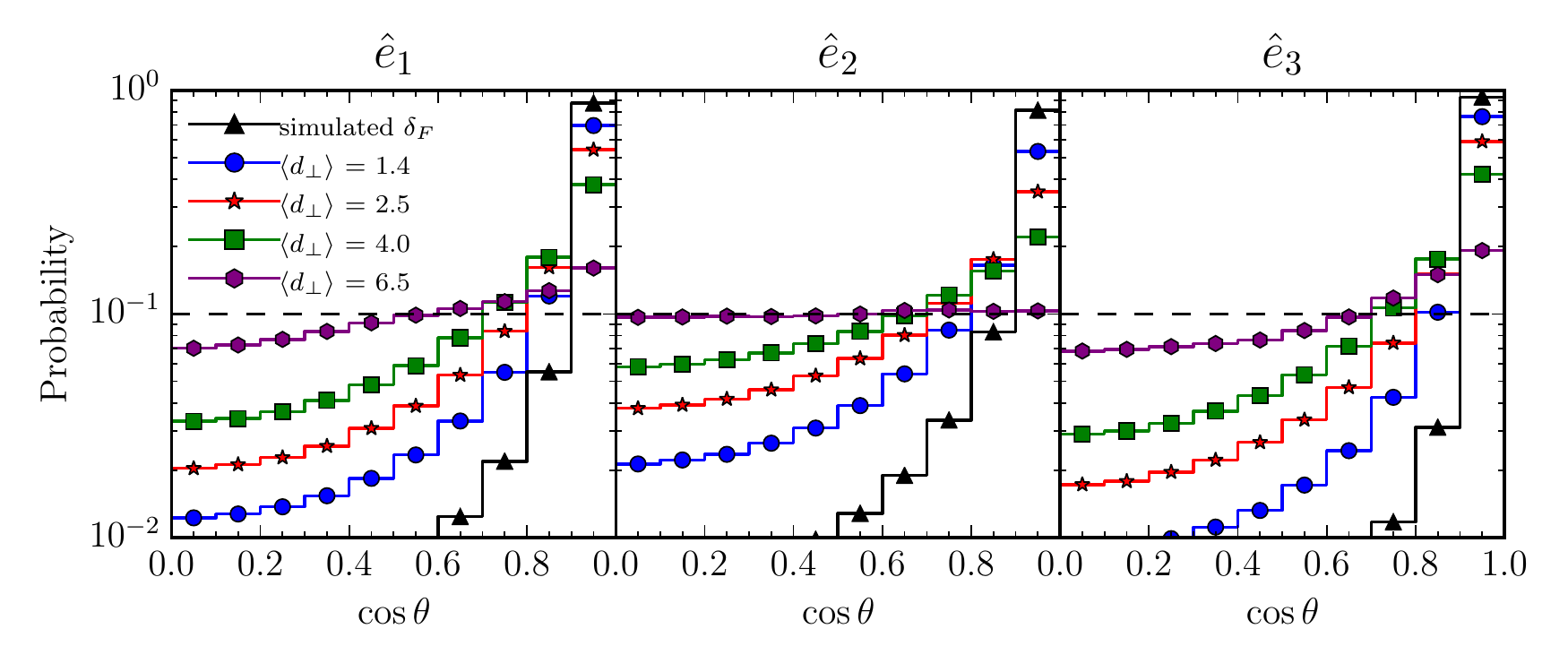,width=20 cm,clip=}}
\caption[]{\small
PDF of the cosine of the misalignment angle between the cosmic web directions
from the matter density field and the cosmic web directions from the simulated
IGM tomography observations.  We also show the misalignment angle between the
matter density cosmic web and the cosmic web computed from the simulated $\delta_F$
with 2 \hMpc{} smoothing, which is the ideal case of a perfect reconstruction
from Ly$\alpha$ forest data.  In all cases the matter density field is smoothed
on 2 \hMpc{} scales.  The horizontal dashed line is the distribution expected
for random alignments.}
\label{fig:evecs_alignments}
\end{figure*}

Figure~\ref{fig:evecs_alignments}
displays the PDF of $\cos{\theta}$,
where $\theta$ is the misalignment angle between the
matter field tidal tensor eigenvectors and the
pseudo-deformation tensor eigenvectors
from the reconstructed IGM maps.
We also compute the misalignment angle PDF between
the matter field tidal tensor eigenvectors and the pseudo-deformation
tensor eigenvectors from the simulated
$\delta_F$ smoothed on 2 \hMpc{} scales, equivalent
to an idealized reconstruction with no instrumental noise
and infinite sightline density.

The mock surveys with \dperp{} $< 5$ \hMpc{} recover
the eigenvectors of the tidal tensor at high significance.
The recovery of the eigenvectors degrades quickly for \dperp{} $> 5$ \hMpc{},
due in part to the mismatch between \dperp{} and the 2 \hMpc{} smoothing
scale of the matter field.
The mean of the cosine of the misalignment angle for [$\hat{e}_1$, $\hat{e}_2$, $\hat{e}_3$]
is [0.874, 0.809, 0.901]
for \dperp{} = 1.4 \hMpc{}, [0.813, 0.716, 0.833] 
for \dperp{} = 2.5 \hMpc{}, [0.736, 0.629, 0.757]
for \dperp{} = 4.0 \hMpc{}, and [0.573, 0.508, 0.600] for \dperp{} = 6.5 \hMpc{}.  Errors on these quantities are $<0.1$\%.
The mean of the cosine of the misalignment
angle between is [0.945, 0.921, 0.967]
using the simulated $\delta_F$ field.  This is
the upper limit of how well Ly$\alpha$ absorption measurements can measure the eigenvectors
of the tidal tensor.

We also compute the mean of the misalignment angle at halo positions only.
For [$\hat{e}_1$, $\hat{e}_2$, $\hat{e}_3$], we find means of [0.891, 0.837, 0.919] for \dperp{} = 1.4 \hMpc{},
[0.815, 0.722, 0.838] for \dperp{} = 2.5 \hMpc{},
[0.734, 0.628, 0.761] for \dperp{} = 4.0 \hMpc{},
[0.571, 0.516, 0.607] for \dperp{} = 6.5 \hMpc{},
and [0.950, 0.929, 0.970] for simulated $\delta_F$.
The differences between these values and the means of $\cos{\theta}$ using the entire grid
are quite modest, although statistically significant.

\begin{figure*}
\hspace*{-1cm}
\centering{\psfig{file=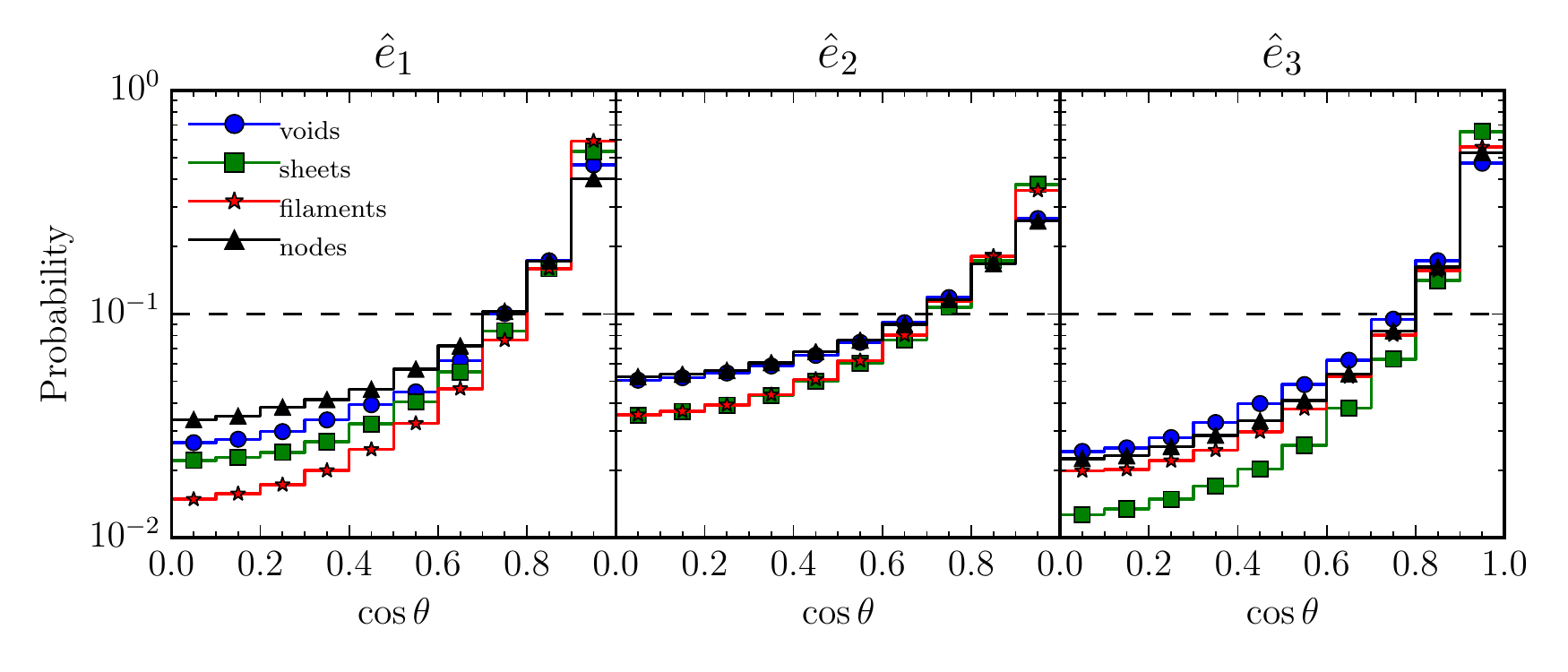,width=20 cm,clip=}}
\caption[]{\small
PDF of the cosine of the misalignment angle between the cosmic web directions
from the matter density field and the cosmic web directions from the simulated
IGM tomography observations, using
the $\langle d_{\perp} \rangle$ = 2.5
\hMpc{} reconstruction and splitting
by classification from the matter map.  All distributions are significantly different according to Kolmogorov-Smirnov
tests.
}
\label{fig:eig_by_web_type}
\end{figure*}

Figure~\ref{fig:eig_by_web_type} shows the quality of cosmic web recovery
as a function of cosmic web type as classified
in the matter map, using the $\langle d_{\perp} \rangle$ = 2.5 \hMpc{}
reconstruction.
Mirroring the results
in Table~\ref{tab:cosmic_web_classification}, $\hat{e}_1$, $\hat{e}_2$, and
$\hat{e}_3$ are recovered worst in nodes.
Despite the relatively high volume overlap for the void recovery,
the tidal tensor eigenvectors are recovered slightly
worse in voids than in anisotropic structures such
as filaments and sheets.  
All three eigenvalues are similar in voids, possibly causing
confusion between perpendicular eigenvalues and leading
to the poorer recovery of the cosmic web in voids.
The recovery of eigenvectors in sheets and filaments are similar,
although $\hat{e}_1$ is recovered better in filaments
while $\hat{e}_3$ is recovered better in sheets.  This is unsurprising
given the connection between the eigenvectors and the geometry
of the cosmic web, as $\hat{e}_1$ in filaments and $\hat{e}_3$
in sheets correspond to inherently anisotropic directions that should
be easier to recover.

\begin{figure*}
\hspace*{-1cm}
\centering{\psfig{file=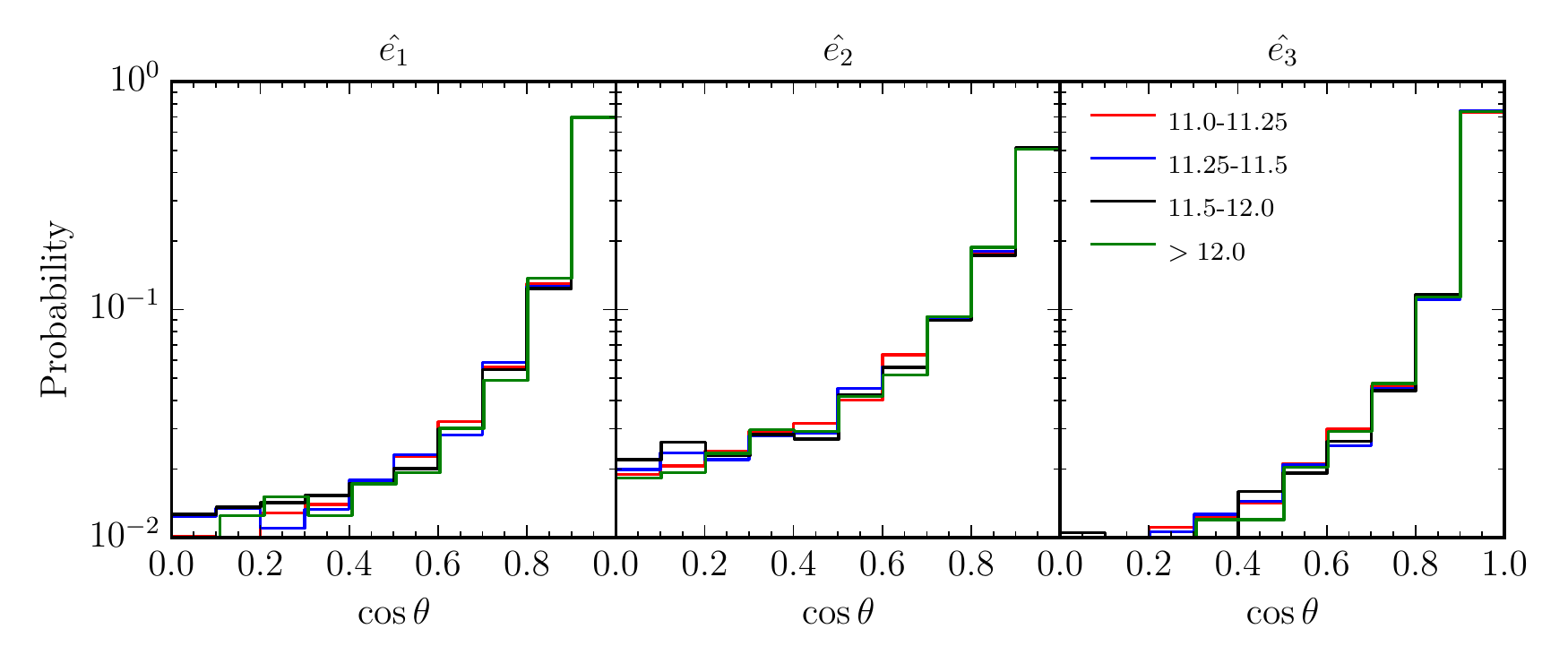,width=20 cm,clip=}}
\caption[]{\small
PDF of the cosine of the misalignment angle between the cosmic web directions
from the matter density field and the cosmic web directions from the simulated
IGM tomography observations, using
the $\langle d_{\perp} \rangle$ = 2.5
\hMpc{} reconstruction.
Misalignment angles are computed at the nearest
grid point for each dark matter halo.  The four
groups are labeled by halo mass in units of
$\log{M_{\odot}}$.
}
\label{fig:e1_by_mass}
\end{figure*}

We additionally test the quality of the reconstruction
as a function of halo mass.
We divide the halo catalog into four bins
and measure the angle between
the eigenvectors in the matter field and the eigenvectors
in the \dperp{}  = 2.5 \hMpc{} reconstruction using the nearest grid
point in redshift space.
The recovery of the eigenvectors is nearly
independent of the halo mass (Figure~\ref{fig:e1_by_mass}).  None of the distributions
are significantly different at the $p=0.05 $ level (2 sigma)
according to Kolmogorov-Smirnov tests, implying
that the eigenvectors can be recovered accurately at galaxy positions
independent of halo mass.

We provide the first assessment of the accuracy of tidal tensor reconstruction
by comparing the tidal tensor eigenvectors from the matter field to the tidal tensor eigenvectors
from realistic mock observations.
We expect that IGM tomography will be better at recovering the tidal tensor eigenvectors
than using the galaxy positions from
existing or upcoming galaxy surveys at the same redshift, such 
as zCOSMOS or PFS;
these will provide much coarser cosmic web maps with galaxy separations of 9 (13) $h^{-1}$ Mpc for PFS (zCOSMOS)
\citep[][]{die+13,takada_pfs}.
For instance, since protocluster identification with zCOSMOS required additional
follow-up spectroscopy \citep{die+15}, the zCOSMOS redshift survey alone may be
insufficient to measure the cosmic web at $z>2$.  In addition, cosmic web maps from
IGM tomography are free from many of the biases that make measuring the cosmic web
from spectroscopic $z \sim 2$ galaxy surveys difficult: these surveys have complex selection
functions \citep[e.g.][]{die+13}, inaccurate and/or biased redshift estimates \citep[e.g.,][]{adelberger:2005,steid+10, rakic:2012}, 
cannot detect close pairs due to slit collisions \citep{wil+15}, and preferentially select
star forming galaxies, which may be biased towards particular regions of the cosmic web \citep{alp15}.
In contrast, IGM tomography sightlines provide an unbiased sampling of the 
foreground cosmic web, and
errors from pixel noise and continuum errors are well-understood, making
a subdominant contribution to errors in the reconstruction.




\subsection{Predictions for galaxy-cosmic web alignment measurements}


Several workers have considered the alignment of galaxies
and the cosmic web in simulations, but thus far observational
studies of the galaxy-cosmic web alignment have generally been
restricted to low redshift ($z\lesssim 0.5$) where sufficiently large
galaxy catalogs exist to measure the cosmic web \citep[][although see \citealt{malavasi16} for recent results at $\langle z\rangle \sim 0.7$]{lp02,le07,temp13,tl13,zhang+13,zhang+15,pah+16}.
We now consider the prospects for a galaxy-cosmic web
alignment study at $z \sim 2$ using the cosmic web
from IGM tomography maps and coeval galaxy samples
from space-based or large ground-based telescopes.

First, we estimate the significance of the measured galaxy-eigenvector
alignment signal as a function of the true strength
of the galaxy-eigenvector alignment.
We describe the PDF of the true galaxy-eigenvector misalignment angle
using Equation~\ref{eqn:align_pdf}, and thus parameterize the galaxy-eigenvector
alignment strength using the deviation of \mcosth{} from 0.5 (i.e. the deviation
from random alignments):
\begin{equation}
\Delta\mcosth \equiv \mcosth - 0.5\end{equation}
The significances
computed from the measured \mcosth{} are similar
to other reasonable choices for estimators, such as using the difference
between the number of aligned and anti-aligned galaxies.
Note that the significance of a model with $- \Delta\langle \cos{\theta}\rangle$
is identical to the significance of a model with $\Delta\langle \cos{\theta}\rangle$
because the spins and eigendirections are headless vectors invariant
under the transformation $\theta \rightarrow \pi - \theta$.
Therefore, we only plot significances as a function of positive $\Delta\langle \cos{\theta}\rangle$.

\begin{figure*}
\hspace*{-1cm}
\centering{\psfig{file=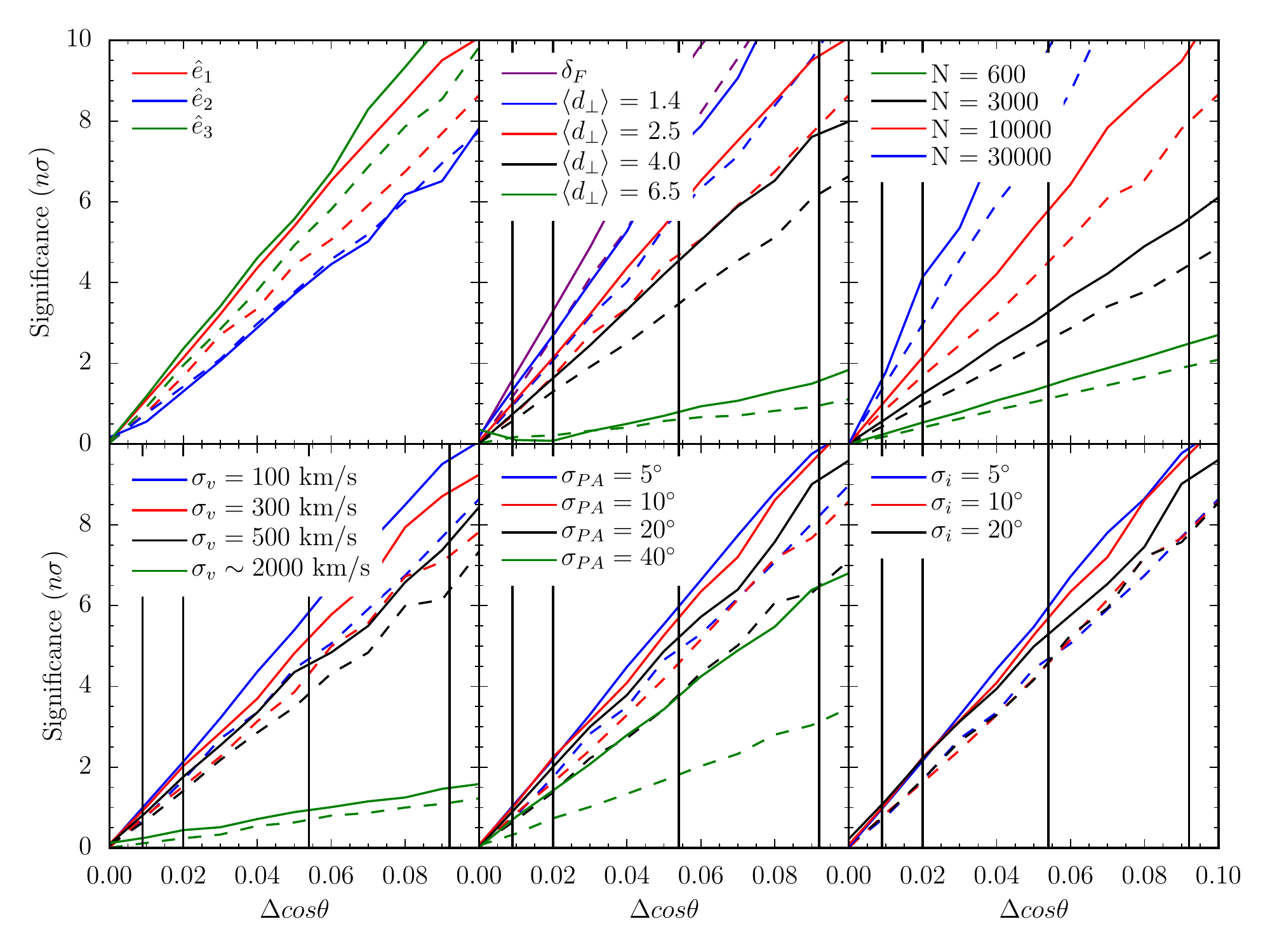,width=20cm,clip=}}
\caption[]{\small
Forecasted significance of the measured galaxy spin-cosmic web alignment
as a function of the true alignment between galaxy spins and the cosmic web.
The $y$-axis indicates the significance of the measurement in units of $\sigma$, while the $x$-axis indicates the deviation of the true alignment
from random.  For all panels except the top left, the galaxy spin axis
is misaligned from \eone{} by an angle drawn from Equation~\ref{eqn:align_pdf} with $\langle \cos{\theta} \rangle = 
\Delta \cos{\theta} + 0.5$.  In the top left panel, the galaxy spin
axis is misaligned from either \eone{}, \etwo{}, or \ethree{}.
For all panels, the solid lines refer to 3D measurements (alignment
between eigenvector and 3D galaxy spin inferred from position angle and ellipticity of the galaxy image) while the dashed
lines refer to 2D measurements (alignment between the eigenvector projected
in the plane of the sky and the galaxy position angle).  
Our fiducial survey has \dperp{} = 2.5 \hMpc{}, 10000 galaxies,
$\sigma_v$ = 100 km s$^{-1}$, $\sigma_{\textrm{PA}} = 10^{\circ}$, and $\sigma_i = 10^{\circ}$.
In each panel, we vary exactly one of these parameters while keeping the
others fixed.
\textit{Top left:} Forecasts for alignments between galaxy spin and
\eone{}, \etwo{} and \ethree{}.  \textit{Top center:} Forecasts
for alignment between galaxy spin and \eone{} for a variety of IGM tomography
surveys with different sightline spacings, including a ``perfect'' survey
where the map is given by $\delta_F$ from the simulation.
Vertical lines indicate different alignment models from simulations, as
defined in Table~\ref{tab:alignment_models}.
\textit{Top right:} forecasts for alignment between galaxy spin
and \eone{} for coeval galaxy samples of different sizes.  
\textit{Bottom left:} Forecasts for alignment between galaxy
spin and \eone{} for different redshift errors.
\textit{Bottom center:} Forecasts for alignment between galaxy spin
and \eone{} for different errors in the galaxy position angle.
\textit{Bottom right:}  Forecasts for alignment between galaxy spin
and \eone{} for different errors in the inclination.  Note that changing
the inclination error does not affect the 2D measurement because it
does not incorporate any information from the inclination.
}
\label{fig:sig2d_3d}
\end{figure*}

In Figure~\ref{fig:sig2d_3d}, we show the forecasted significance
of the alignment signal between galaxies and \eone{}, 
as a function of the sightline spacing, number of galaxies, expected
redshift error, and errors on the galaxy position angle and inclination.
Our fiducial values for these quantities are \dperp{} = 2.5 \hMpc{},
10000 coeval galaxies, redshift errors of 100 km s$^{-1}$, and position
angle and inclination errors of 10$^{\circ}$.  The 1-$\sigma$ error on
$\Delta\mcosth$ of $\sim 0.01$
for the fiducial measurement of alignment with \eone{}
is somewhat larger than 1-$\sigma$ errors from alignment measurements
using low-redshift galaxies \citep[$\sim 0.005$ for similar galaxy sample sizes
in][]{temp13,tl13,pah+16}.

\begin{deluxetable*}{l p{7cm} l}
\tablecolumns{3}
\tablecaption{\label{tab:alignment_models} Summary of alignment models}
\tablehead{
\colhead{$\Delta \langle\cos{\theta}\rangle$} &
\colhead{Description} &
\colhead{Reference}
}
\startdata
0.009 & $z \sim 1.2$ alignments between \eone{} and galaxy spin from HorizonAGN &\citet{codis_IA} \\[1.5em]
0.020 & $z \sim 2.3$ alignments between filaments  and galaxy spins from HorizonAGN
 & \citet{dub+14} \\[1.5em]
0.054 & $z \sim 1.05$ alignment between \eone{} and halo shape (N-body), plus misalignment between
halo shape and galaxy shape, alignments extrapolated to $z \sim 2.4$ 
and $M_h \sim 10^{12} M_{\odot}$ assuming alignments constant as a function of $M_h$ &
\citet{hahn+07_evolution,oku+09} \\[1.5em]
0.092 & $z \sim 1.05$ alignment between \eone{} and halo shape (N-body), plus misalignment between
halo shape and galaxy shape, alignments extrapolated to $z \sim 2.4$ 
and $M_h \sim 10^{12} M_{\odot}$ assuming alignments constant as a function of $M_h/M_{nl}$ &
\citet{hahn+07_evolution,oku+09} \\
\enddata
\tablecomments{Alignment models considered in Figure~\ref{fig:sig2d_3d}.  See text for details. Each alignment model is parameterized using Equation~\ref{eqn:align_pdf} with \mcosth{} = 0.5 + $\Delta \cos{\theta}$.}
\end{deluxetable*}

Figure~\ref{fig:sig2d_3d} shows the significance of alignment
 measurements between galaxy spins and \eone{}, \etwo{}, and \ethree{}
 as a function of the true alignment strength between galaxy spin
 and \eone{}, \etwo{} and \ethree{}.  Consistent with Figure~\ref{fig:evecs_alignments}, alignments
 with \eone{} and \ethree{} can be detected at similar levels of significance,
 while alignments with \etwo{} will be detected with somewhat lower significance.
 We note that \etwo{} alignments are detected at similar significance for both 2D and 3D measurements.
 We attribute this to redshift-space distortions in the map, which cause \etwo{} to lie preferentially
 in the plane of the sky, because the directions of maximal and minimal compression (\eone{} and \ethree{})
 are biased towards the line of sight due to compression and rarefaction from redshift-space distortions.
 Measuring alignments between \etwo{} from the real-space density field and
 \etwo{} from skewers generated using the real-space $\tau$ yields a similar reduction in
 the 2D \etwo{} alignment signal as for the 2D \eone{} and \ethree{} alignment signals.

 Tidal torque theory predicts that galaxy spin is aligned with \etwo{}.
 Using an analytic quadratic alignment model \citep{lp00},
 \citet{le07} derive a PDF for the misalignment angle as a function of $c$, a
 correlation parameter ranging from 0 (random alignments) to 1 (perfect spin-shear alignments).
 In the limit of small $\Delta\mcosth$, this PDF can be well approximated by
 Equation~\ref{eqn:align_pdf}.  For our fiducial \etwo{} alignment measurement we find a 1-$\sigma$
 error on $\Delta\mcosth \sim 0.015$, which translates to a 1-$\sigma$ error on $c$ $\sim 0.08$.
 Previous measurements at low redshift
 have achieved lower error bars: \citet{lp02} find $c = 0.28 \pm 0.07$
 \footnote{\citet{lp02} expresses their result in terms of $a_T = (3/5)c$. See
\citet{lp01} for the difference between $a_T$ and $c$.},
\citet{le07} find $c = 0.08 \pm 0.01$,
 and \citet{lp07} find $c = 0.0 \pm 0.05$ for red galaxies and
 $c = 0.33 \pm 0.07$ for blue galaxies.  If $c \sim 0.2-0.3$
 at $z \sim 2.4$,
IGM tomography surveys with \dperp{} = 2.5 \hMpc{}
and 10000 coeval galaxies will measure this alignment
at $\sim3\sigma$.  However, nonlinear evolution
 is expected to decrease the alignment between spin and \etwo{}
 over time \citep{por02},
 leading to a larger value of $c$ at high redshift than low redshift.  Regardless,
 these results suggest that the combination of galaxy surveys at low redshift
 and IGM tomography surveys at high redshift may be able to constrain
 the redshift evolution of the alignment and thus provide a more rigorous
 test of tidal torque theory than current studies.

Figure~\ref{fig:sig2d_3d} shows the importance of varying different parameters
of the IGM and galaxy observations.
Varying the sightline spacing from $\dperp = 1.4\,\hMpc$ to $\dperp = 4\,\hMpc$ is equivalent to decreasing
the number of sightlines by an order of magnitude.  Thus, comparison of the
top center and top right panels of Figure~\ref{fig:sig2d_3d} shows that close to the
fiducial sightline spacing of 2.5 \hMpc{}, increasing the number of coeval galaxies
is more important than increasing the number of background Ly$\alpha$ forest sightlines.
However, the measured significance of the signal drops dramatically
for sightline spacings $\dperp> 4$ \hMpc{}.
This suggests that a wide-field survey such as PFS is preferable for measuring
galaxy-cosmic web alignments, as it would be best positioned to deliver a large
coeval galaxy sample, while the coarser sightline spacing of 4 \hMpc{} only modestly
lowers the significance compared to the 2.5 \hMpc{} for CLAMATO.  However, the poor performance of the \dperp{}  = 6.5
\hMpc{} survey
suggests that constraints on galaxy-cosmic web alignments from PFS will require a supplementary tomography
component to achieve the necessary sightline spacing.

The impact of spectroscopic redshift errors are quite modest, as Figure~\ref{fig:sig2d_3d}
shows little difference between a sample with redshift errors $\sim 100$ km s$^{-1}$
(characteristic of redshifts measured from rest-frame optical nebular emission
lines) and a sample with redshift errors
$\sim 300$ km s$^{-1}$ (characteristic of redshifts measured from rest-frame UV absorption lines).
The impact on the significance is quite small even if the redshifts were
from emission lines in a grism spectrum ($\sigma_v \sim 500$ km s$^{-1}$)
However, a redshift error of $\sim 2000$ km s$^{-1}$ from the randomized sample leads
to a drastic reduction in the constraining power of the survey.  Since this redshift error
is already optimistic for photometric redshifts, we conclude that spectroscopic redshifts are essential
to measure spin-cosmic web alignments.

Varying the fiducial position angle and inclination errors by a factor of two
makes relatively little difference for measuring the alignment signal. 
We also test a model with a relatively large PA error of 40$^{\circ}$, which
may be more representative of the shape errors in a ground-based survey.
In this case, much more of the constraining power is coming from the inclination.
If the survey is also unable to recover the inclinations (due to large
uncertainties in measuring the axis ratio),
the significance of the alignment signal drops dramatically.
Therefore, reasonably precise estimates of the position error
($\sigma_{\textrm{PA}} \lesssim 30^{\circ}$) will be necessary to measure
alignments between galaxy spin and the cosmic web.


We also explore the possibilities for constraining
alignment models based on results from simulations.  Simulations have measured
a broad variety of alignments (e.g. between halo/galaxy spins/shapes and filaments/\eone{})
across a wide range of redshifts.  In general, simulations have found stronger alignments
between halo shapes and the cosmic web than between halo spins and the cosmic web
\citep{hahn+07_evolution}.
While we have framed the above discussion in terms of galaxy spin alignments,
ultimately we are measuring the major axis of the galaxy image, which may be set
by either the spin or the shape of the galaxy and ultimately the dark matter halo.
Therefore, we consider alignment models from both galaxy spin and halo shape.

We use the simulation results of \citet{hahn+07_evolution} to create an alignment model
based on halo shape.  
\citet{hahn+07_evolution} studies the alignment
between halo shape and \eone{} at $z$ = 0, 0.49, and 1.05.
After accounting for the mass dependence by scaling by the mass scale
of nonlinear collapse $M_*(z)$
they find no additional redshift dependence.

We extrapolate the results of \citet{hahn+07_evolution} to $z \sim 2.3$
to create an alignment model.
\citet{hahn+07_evolution} presents the mass
dependence of their result using $M_h/M_*$ where $M_*$ is the mass
for which a 1-$\sigma$ fluctuation reaches the threshold for spherical
collapse, $\delta_c = 1.686$.  At $z \sim 2.3$, the typical halo ($M_h \sim 10^{12} M_{\odot}$)
in the galaxy sample has $M \gtrsim 100 M_*$, leading to an extremely
strong alignment signal, median $\cos{\theta}$ = 0.78 (\mcosth{} = 0.73
assuming the PDF follows Equation~\ref{eqn:align_pdf}).
More conservatively, it is possible that above $z \sim 1.05$ the alignment
signal is constant with $M_h$ rather than $M_h/M_*$.  In this case, the
typical halo detected in our galaxy sample would have $M \sim 10 M_{*,z=1.05}$,
leading to a slightly weaker median $\cos{\theta}$ = 0.7
(\mcosth{} = 0.64) between \eone{} and halo shape.
To translate from halo shape alignments to galaxy shape alignments,
we use the halo-galaxy misalignment model of \citet{oku+09},
a Gaussian distribution of halo-galaxy misalignment angles with dispersion 35$^{\circ}$.
Therefore, we find 
$\Delta$\mcosth{} = 0.092 (0.054) for the \citet{hahn+07_evolution}
model using $M_h/M_*$ ($M_h$) to determine alignments.

In contrast to the shape measurements, several workers have measured alignments
between galaxy spins and the cosmic web using hydrodynamic simulations of the galaxies.
These include \citet{codis_IA}, who measure alignments between galaxy spin and \eone{}
 at $z \sim 1.2$, finding $\Delta$\mcosth{} = 0.009.  Similarly, \citet{dub+14}
 measure alignments between galaxy spin and filaments for galaxies between
 $z \sim 1.2$ and $z \sim 3$ and find $\Delta$\mcosth{} = 0.02.  These alignments
 are similar to alignments inferred from simulations of halo spin-cosmic web alignments
 \citep{hahn+07_evolution,trow+13} with a galaxy-halo spin misalignment following
 \citet{bett12}.  The spin-web alignment is considerably weaker than the shape-web
 alignment, consistent with results using only dark matter halos \citep{hahn+07_evolution} and possibly
 related to stronger shape-cosmic web alignments for early-type than late-type
 galaxies observed at low redshift \citep{tl13,pah+16} and the
 considerably stronger intrinsic alignments observed for early-type
  than late-type galaxies \citep{joa+13_1,mand+06}.
 
 Figure~\ref{fig:sig2d_3d} shows that IGM tomography surveys in tandem with wide-field
 galaxy surveys would be able to detect or rule out alignment models based on halo shape
 (e.g. from \citet{hahn+07_evolution} at high significance.  
 Even the combination of the CLAMATO survey and zCOSMOS galaxy survey (with
 $N_{\textrm{gal}} \sim 600$ over the $V\sim 10^6\,h^{-3}\mathrm{Mpc}^3$ CLAMATO volume at $2.1<z<2.5$) 
 will be sufficient
 to detect or rule out the most aggressive alignment models based on halo shape at $\sim$2-3 $\sigma$.  However, more realistic models with smaller alignments
 will require an order of magnitude more coeval galaxies for detection
 at a similar level.  Figure~\ref{fig:sig2d_3d} also shows that ground-based
 imaging should be sufficient to measure galaxy spins as long as position angles
 can be measured with an error $\lesssim 20^{\circ}$.
 

Additional measurements besides spin-eigendirection correlations may yield
further independent information.  For instance, IGM tomography will be able
to identify a large number of voids \citep{stark_voids}, allowing measurement of
the void-spin correlation \citep[e.g.][]{tru+06,sw+09,var+12}.  We can also use
the eigenvalues to partition our map into voids, sheets, filaments, and nodes
\citep{lee_white+16} and measure spin-\eone{} alignments only in filaments, or 
spin-\ethree{} alignments only in sheets, where they may be strongest
\citep[e.g.][]{hahn+07_evolution}.

\section{Conclusions}
Intrinsic alignments between galaxies and the underlying cosmic web 
have been predicted from both DM-only and hydrodynamical simulations,
several of which predict increasing alignment strength at higher redshift ($z\gtrsim 1$).
At these redshifts, it becomes increasingly expensive to obtain spectroscopic redshifts
with sufficiently high number densities to trace the cosmic web. 
Recently, tomographic reconstruction of the IGM as traced by high area densities
of Lyman-$\alpha$ forest sightlines has emerged as a promising method to map the 
cosmic web at $z\sim 2-3$. 

In this paper, we studied the feasibility of IGM tomographic surveys, in conjunction
with coeval galaxy redshift samples with measured structural parameters, 
to place constraints on galaxy-cosmic web alignments at $z\sim 2.5$.
Using detailed hydrodynamical simulations based on the Nyx code, we first generated realistic
mock data sets reflecting both ongoing and future IGM tomography surveys. 
The galaxy spin or shape distributions were `painted on' with respect to the underlying
matter tidal tensor field using a simple alignment model 
parameterized by $\Delta\langle\cos{\theta}\rangle$, i.e.\ the non-random excess alignment of the
galaxies with respect to the eigenvectors of the matter tidal tensor.
Future studies of galaxy-cosmic web alignments at $z \sim 2$ will benefit
greatly from simulations combining both realistic IGM physics and
galaxy formation (e.g. future versions of the Nyx simulations used in this paper).

First, we showed that IGM tomography with sightline separations of $\dperp \leq 5\,\hMpc$ 
should be able to recover the eigenvectors
of the tidal tensor, \eone{}, \etwo{}, and \ethree{}, as determined from the large-scale distribution
of matter in the universe (smoothed on $2\,\hMpc$ scales).  The mean dot products
between the eigenvectors as determined by the matter field and the eigenvectors from a mock
observation with sightline spacing 2.5
\hMpc{}, 
similar to the ongoing CLAMATO survey, are [0.815, 0.722, 0.838] for [\eone{}, \etwo{}, \ethree{}].
This builds on our previous result
showing that IGM tomography can recover eigenvalue cosmic web classifications
with a fidelity similar to $z \lesssim 0.7$ surveys \citep{lee_white+16}.

We then compared the eigenvectors recovered from the IGM tomography with the spins or shapes
in coeval galaxy samples
as a function of the alignment strength  $\Delta\langle\cos{\theta}\rangle$, and also considered
the effect of uncertainties in the measurement of the galaxy position angles, inclinations, and
redshift estimation. 
The largest factor in our ability to constrain the galaxy-cosmic web alignments is the size of the galaxy sample.
Assuming a fiducial mean sightline separation of $\dperp =2.5\,\hMpc$, redshift errors of $\sigma_v =500\,\mathrm{km s^{-1}}$, as well as errors of $\sigma_{\textrm{PA}}=10\deg$
and $\sigma_i =10\deg$ in the galaxy position angles and inclinations, respectively, we find that
the ongoing CLAMATO Survey on the Keck-I Telescope\footnote{Comoving volume of $V\sim 10^6\,h^{-3}\,\mathrm{Mpc}^3$
over $2.1<z<2.5$ within the central square degree of the COSMOS Field.}, in conjunction with $\sim 600$
coeval galaxies from zCOSMOS-Deep and other spectroscopic surveys, should be able to place $\sim 3\sigma$
limits on the most extreme alignment models with $\Delta\langle\cos\theta\rangle\sim 0.1$ within the next few years.
For most alignment models with $\Delta\langle\cos\theta\rangle<0.05$, however, coeval samples of at least 
several thousand galaxies with spectroscopic redshifts would be needed to make a $\sim 3-4\sigma$ detection.
These results are not very sensitive to the mean sightline separation of
the IGM tomography survey so long as $\dperp \lesssim 5\,\hMpc$,
nor on the accuracy of the galaxy structural parameters, although space-based imaging
or very good quality ground-based imaging ($<0.5$ arcsec seeing) would be desirable for the latter.
We find that photometric redshifts are insufficient for this purpose as the redshift errors are far too large.

Since the primary limitation for this alignment measurement is the size of available galaxy 
redshift samples at $z\sim 2.5$, a relatively wide/shallow strategy would be optimal:
at fixed survey magnitude, the galaxy sample size $N_{\textrm{gal}}$ scales linearly with telescope time
by expanding survey area. On the other hand, increasing
$N_{\textrm{gal}}$ by increasing survey depth within a small survey 
area would require  exponential increases of telescope time.
Since the tidal tensor eigenvector recovery does not degrade
much with slightly coarser tomographic reconstructions 
relative to the fiducial $2.5\,\hMpc$ sightline spacing
in CLAMATO, this argues that
near-future wide-field instruments, i.e.\ Subaru PFS, can cover much larger
areas than CLAMATO with the concomitantly larger $N_{\textrm{gal}}$ for significantly 
improved spin-cosmic web or shape-cosmic web constraints. 
However, we did find a mean spacing of 
$\dperp < 5\,\hMpc$ is required for the Ly$\alpha$ forest sightlines,
above which the eigenvector recovery 
degrades considerably. Since the $2<z<3$ LBG component currently planned for the 
$\sim 20-30$ sq deg PFS Galaxy Evolution Survey leads to a ``free'' IGM tomographic map 
with $\dperp \sim 6.5\,\hMpc$, we advocate supplemental PFS spectroscopy to boost
the sightline sampling to $\dperp\approx 4\,\hMpc$. Based on 
the calculations of \citet{lee_obs_req}, this should require
$\sim5-6$hrs of additional exposure time per field, or 
$\sim 20$ nights over $\sim 25$ sq deg (including weather/seeing overheads).
Such a program, along with the $\sim 10,000$ coeval galaxies
also from PFS, should allow $3\sigma$ limits on
alignments down to $\Delta\langle\cos\theta\rangle\approx0.03$.
Constraints on even smaller $\Delta\langle\cos\theta\rangle$,
at the levels predicted by, e.g. \citet{codis_IA}, would require
even more ambitious surveys. However, it is conceivable that a new
generation of massively-multiplexed wide-field spectrographs on 
$>$10m-class telescopes could be available by the early 2030s
\citep{mcconnachie:2016,dod+16,najita:2016}, in time to provide
priors on the intrinsic alignment systematics for the final LSST
tomographic weak lensing analyses.

\acknowledgments
We thank Peter Nugent, Miguel Aragon-Calvo,
Nadia Zakamska, Joanne Cohn, and
Sedona Price for useful discussions and comments.
K.G.L. acknowledges support for this
work by NASA through Hubble Fellowship grant HF2-51361
awarded by the Space Telescope Science Institute, which is
operated by the Association of Universities for Research in
Astronomy, Inc., for NASA, under contract NAS5-26555.
ZL and AK were in part supported by the Scientific Discovery through Advanced Computing (SciDAC) program funded by U.S. Department of Energy Office of Advanced Scientific Computing Research and the Office of High Energy Physics.
Calculations presented in this paper used resources of the National Energy Research Scientific Computing Center (NERSC), which is supported by the Office of Science of the U.S.~Department of Energy under Contract No.~DE-AC02-05CH11231.

\bibliographystyle{yahapj}
\bibliography{citations.bib}


\end{document}